\documentclass[12pt]{article} 
\usepackage{graphicx} 
\def \b{{\cal B}} 
\def \bea{\begin{eqnarray}} 
\def \beq{\begin{equation}}

\def \eea{\end{eqnarray}} 
\def \eeq{\end{equation}} 
\def \ite{{\it et al.}}

\def \s{\sqrt{2}} 
\def \st{\sqrt{3}} 
\def \sx{\sqrt{6}} 
\def\lsim{\mathrel{\rlap{\lower3pt\hbox{$\sim$}}\raise2pt\hbox{$<$}}}
\def\gsim{\mathrel{\rlap{\lower3pt\hbox{$\sim$}}\raise2pt\hbox{$>$}}}

\renewcommand{\thetable}{\Roman{table}}
\begin{document} 
\begin{flushright}
TECHNION-PH-2010-19 \\
EFI 10-31 \\
arXiv:1012.5098 [hep-ph] \\ 
December 2010 \\ 
\end{flushright} 
\renewcommand{\thesection}{\Roman{section}} 
\renewcommand{\thetable}{\Roman{table}}
\centerline{\bf RATIOS OF HEAVY HADRON SEMILEPTONIC DECAY RATES}
\medskip
\centerline{Michael Gronau} 
\centerline{\it Physics Department, Technion, Haifa 32000, Israel.}
\medskip 
\centerline{Jonathan L. Rosner} 
\centerline{\it Enrico Fermi Institute and Department of Physics,
  University of Chicago} 
\centerline{\it Chicago, IL 60637, U.S.A.} 
\bigskip 

\begin{quote}
Ratios of charmed meson and baryon semileptonic decay rates appear to be
satisfactorily described by considering only the lowest-lying (S-wave)
hadronic final states and assuming the kinematic factor describing phase
space suppression is the same as that for free quarks.  For example, the
rate for $D_s$ semileptonic decay is known to be $(17.0 \pm 5.3)\%$ lower
than those for $D^0$ or $D^+$, and the model accounts for this difference.
When applied to hadrons containing $b$ quarks, this method implies that the
$B_s$ semileptonic decay rate is about $1\%$ higher
than that of the nonstrange $B$ mesons.  This small difference thus suggests
surprisingly good local quark-hadron duality for $B$ semileptonic decays,
complementing the expectation based on inclusive quark-hadron duality that
these differences in rates should not exceed a few tenths of a percent.  For
$\Lambda_b$ semileptonic decay, however, the inclusive rate is predicted to
be about 13\% greater than that of the nonstrange $B$ mesons.  This value,
representing a considerable departure from a calculation using a heavy quark 
expansion, is close to the corresponding experimental ratio $\Gamma(\Lambda_b)/
\bar \Gamma(B) = 1.13 \pm 0.03$ of total decay rates. 
\end{quote}

\leftline{\qquad PACS codes: 13.25.Fc, 13.20.He, 13.30.Ce}

\section{Introduction}

An early prediction for charmed meson decays \cite{Pais:1977nn}, based on
isospin symmetry, was the equality of Cabibbo-favored $D^0$ and $D^+$
semileptonic decay rates, borne out by experiment within errors \cite{PDG}.  On
the other hand, the $D_s$ semileptonic decay rate is now known to be about
$(17.0 \pm 5.3)\%$ lower than the average of the $D^0$ and $D^+$ rates.  This
difference not only sheds light on strong-interaction dynamics, but can serve
as a useful calibration when tagging $D_s$ decays.  The corresponding ratio of
strange and non-strange $B$ meson semileptonic decay rates is much less well
known,
and there is only fragmentary information on $\Lambda_b$ semileptonic decays.

In the present paper we briefly review what heavy-quark symmetry has to say
about the ratios of semileptonic decay rates of various hadrons containing
heavy ($c$ and $b$) quarks (Sec.\ \ref{sec:hq}).  We then introduce an
effective-quark method for comparing decays by means of the kinematic factor
that characterizes $\mu^- \to e^- \bar \nu_e \nu_\mu$
when the electron mass is not neglected (Sec.\
\ref{sec:eq}).  This method is shown in Sec.\ \ref{sec:c} to reproduce the
relative suppression of the $D_s$ semileptonic rate and the apparent
enhancement of the $\Lambda_c$ semileptonic rate (still not very precisely
measured).  When applied to $B$, $B_s$, and $\Lambda_b$ decays (Sec.\
\ref{sec:b}), it leads to the prediction of relative {\it enhancements} of the
$B_s$ and $\Lambda_b$ semileptonic rates by $\sim 1\%$ and $\sim 13\%$,
respectively, with respect to those of the nonstrange $B$ mesons.  Verification
of these predictions would be surprisingly good evidence for local
quark-hadron duality for mesons, complementing the expectation based on the
operator product expansion \cite{Bigi:2001ys} that differences in semileptonic
decay rates of mesons containing $b$ quarks should not exceed a few tenths 
of a percent.  For baryons the large deviation from unity is similar to that
observed in total decay rates.  Prospects for checking these
predictions, and a summary, are contained in Sec.\ \ref{sec:sum}.

\section{Expectations from heavy-quark symmetry \label{sec:hq}}

The relation between semileptonic decays of free quarks and those of hadrons
is based on the notion of quark-hadron duality, whose origins and concepts
are well-described in the review of Ref.\ \cite{Bigi:2001ys}.  The
corrections to a free-quark picture may be framed in terms of an
operator-product expansion involving terms proportional to inverse powers
of the mass $m_Q$ of the decaying quark and to powers of the strong coupling
constant $\alpha_S$ \cite{Chay:1990da}.  For an early discussion of the
magnitude of such terms, see Ref.\ \cite{Bigi:1992su}.

Corrections of ${\cal O}(1/m_Q)$ to the free-quark picture were proposed in
Refs.\ \cite{Isgur:1998ws,Isgur:1999cd}.  The absence of such terms was shown
in Ref.\ \cite{Bigi:2001ys} to involve non-trivial cancellations.  Terms of
${\cal O}(m_s \Lambda_{QCD}/m_Q^2)$ can affect the semileptonic rates.  One can
scale the observed difference of $(17.0 \pm 5.3)\%$ between the strange
and nonstrange $D$ meson semileptonic decay rates by a factor of $(m_c/m_b)^2
\sim 0.1$ to estimate a difference of no more than a percent or two for
strange and nonstrange $B$ mesons.

Non-perturbative corrections which violate flavor symmetry have been estimated
not to exceed half a percent for $b$ semileptonic decays \cite{Bigi:2001ys}.  A
more recent investigation \cite{Bauer:2004ve} finds no evidence for violation
of quark-hadron duality in inclusive $b \to c$ decays, and concludes that
in the limit of $m_b \gg \Lambda_{QCD}$ inclusive $B$ decay rates are equal
to the $b$ quark decay rates.  A similar conclusion is reached by Kowalewski
and Mannel in a mini-review of determinations of the Cabibbo-Kobayashi-Maskawa
(CKM) matrix elements $V_{cb}$ and $V_{ub}$ in Ref.\ \cite{PDG}.

The above arguments apply to inclusive semileptonic decays.  These decays
populate the lowest-lying hadronic final states, consisting of the lowest
S-wave $q \bar q$ or $qqq$ levels, but such final states do not saturate the
total semileptonic widths.  The P-wave levels are excited to a degree, and
there may also be contributions from non-resonant hadronic continuum.
This leads to non-trivial form factors for excitation of the lowest levels.

The question then arises:  To what degree are the conclusions of quark-hadron
duality mirrored in the ratios of decay rates to the lowest-lying levels?
In the present discussion we offer a simplified model predicting the ratios
of semileptonic decay rates just on the basis of transitions to the
lowest-lying levels.  We are encouraged by the fact that this model reproduces
the relative suppression of the $D_s$ semileptonic decay rate and the
(less-well-measured) apparent enhancement of the $\Lambda_c$ semileptonic
decay rate correctly.  We find that such a model predicts effects of order
one percent for mesons containing a $b$ quark, in accord with the
na\"{\i}ve scaling arguments presented above.  For baryons, however, we find
roughly the same enhancement of semileptonic decay rate with respect to those
of the nonstrange $B$ mesons, roughly 13\%, that is seen in total decay rates
\cite{PDG}.

\section{Calculations at effective quark level \label{sec:eq}}

Semileptonic decays of charm and beauty mesons involve
final states consisting largely, though not exclusively, of the lowest-lying
pseudoscalar and vector mesons.  Cabibbo-favored charm decays and 
Cabibbo-Kobayashi-Maskawa (CKM)-favored beauty decays are noted 
in Table \ref{tab:decays}, along with kinematic factors defined by
\beq
\label{eqn:f}
f(x) \equiv 1 - 8x + 8 x^3 - x^4 + 12 x^2 \ln(1/x)~,~~~~~~x\equiv (M_f/M_i)^2~,
\eeq
where $M_i$ and $M_f$ are masses of decay and daughter mesons.  For decays
related by isospin, we have quoted charge-averaged initial and final masses.
The kinematic factor $f(x)$ was applied to semileptonic charm decays, for
example, in Refs.\ \cite{Gaillard:1974mw,Gershtein:1976aq}.

\begin{table}
\caption{Cabibbo-favored semileptonic decays of charmed hadrons and CKM-favored
semileptonic decays of beauty hadrons to lowest-lying S-wave states.  $M_i$ and
$M_f$ are masses of initial and final hadronic states based on \cite{PDG}, $x
\equiv (M_f/M_i)^2$, and $f(x)$ is defined in Eq.\ (\ref{eqn:f}).
\label{tab:decays}}
\begin{center}
\begin{tabular}{c c c c c} \hline \hline
Decay & $M_i$ (MeV/$c^2$) & $M_f$ (MeV/$c^2$) & $x$ & $f(x)$ \\ \hline
$D \to \bar K \ell^+ \nu_\ell^{~a}$ & 1867.22 & 495.65 & 0.07046 & 0.5971 \\
$D \to \bar K^* \ell^+ \nu_\ell^{~a}$ & 1867.22 & 893.80 & 0.22913 & 0.1887 \\
$D_s^+ \to \eta \ell^+ \nu_\ell$ & 1968.47 & 547.85 & 0.07746 & 0.5682 \\
$D_s^+ \to \eta' \ell^+ \nu_\ell$ & 1968.47 & 957.78 & 0.23674 & 0.1781 \\
$D_s^+ \to \phi \ell^+ \nu_\ell$ & 1968.47 & 1019.46 & 0.26821 & 0.1395 \\
$\Lambda_c^+ \to \Lambda \ell^+ \nu_\ell$ & 2286.46 & 1115.68 & 0.23810 &
0.1763 \\
$\bar B \to D \ell^- \bar \nu_\ell^{~a}$ & 5279.34 & 1867.22 & 0.12509 & 0.4050
\\
$\bar B \to D^* \ell^- \bar \nu_\ell^{~a}$ & 5279.34 & 2008.61 & 0.14475
 & 0.3518 \\
$\bar B_s \to D_s^+ \ell^- \bar \nu_\ell$ & 5366.3 & 1968.47 & 0.13456 & 0.3785
\\
$\bar B_s \to D_s^{*+} \ell^- \bar \nu_\ell$ & 5366.3 & 2112.3 & 0.15494 &
 0.3268 \\
$\Lambda_b \to \Lambda_c \ell^- \bar \nu_\ell$ & 5620.2 & 2286.46 &
 0.16551 & 0.3027 \\ \hline \hline
\end{tabular}
\end{center}
\leftline{$^a$ Charge-averaged masses.}
\end{table}

We model the effects of initial- and final-state mass differences by assuming
that decays are characterized by the kinematic factor $f(x)$.  Strictly
speaking, this factor applies to the decay of a fermion into three fermions,
e.g., $\mu^- \to e^- \bar \nu_e \nu_\mu$, in which one of the final fermions
(the $e$ in this example) has non-zero mass.  We then characterize the decay
rate by assuming it is proportional to $2J_f + 1$, where $J_f=0,1$ is the spin
of the final state, assumed to be either the ground-state pseudoscalar meson or
the ground-state vector meson.  These are the states listed in Table
\ref{tab:decays}.  The effective kinematic factor $\bar f$ is then the
spin-weighted average of that for a pseudoscalar final state (weight 1/4)
and a vector final state (weight 3/4).  Thus, for example,
\beq\label{eq:fD}
\bar f(D) = \frac{1}{4} f\left( \left[ \frac{\bar M(K)}{\bar M(D)} \right]^2
\right) + \frac{3}{4} f\left( \left[ \frac{\bar M(K^*)}{\bar M(D)} \right]^2
\right) = \frac{1}{4}(0.5971) + \frac{3}{4}(0.1887) = 0.2908~.
\eeq
Here $\bar M$ refers to the charge-averaged masses quoted in Table
\ref{tab:decays}. 

The lowest-lying baryons containing a heavy quark $Q$ are of the form
$\Lambda_Q = Q[ud]$, where the brackets denote a $ud$ pair of spin zero and
isospin zero.  In the approximation we are making, the only final states
considered when $Q \to Q' \ell \nu_\ell$ are those in which the final quarks
are in an S-wave and the $ud$ pair continues to have zero spin and isospin.
Thus, we consider only $\Lambda_c \to \Lambda \ell^+ \nu_\ell$ and $\Lambda_b
\to \Lambda_c \ell^- \bar \nu_\ell$.

This model is an oversimplification when applied to the calculation of
actual decay rates.  For example, it neglects form factors for decays into the
lowest-lying hadrons and important branching fractions to higher-lying hadronic
final states, and does not accurately represent the vector-to-pseudoscalar
ratio of semileptonic rates.  Nevertheless, we may expect it to be useful for
calculating {\it ratios} of semileptonic decay rates, as the kinematic
differences between semileptonic decays of different hadrons containing the
same heavy quark are mirrored to some extent in decays to higher-lying states.
We expect the use of free quark kinematics with quark masses replaced by hadron
masses and the averaging over the spins of the final states (for meson decays)
to mimic the effects of confinement.  Our ``cartoon'' version of
quark-hadron duality ultimately should be replaced by an approach based on the
operator product expansion and heavy quark effective theory. 
We shall compare a few of our predictions with those of the latter approach.

\section{Application to semileptonic charm decays \label{sec:c}}

We first check the equality of semileptonic $D^0$ and $D^+$ decay
rates as predicted for Cabibbo-favored decays.  We assume $(e,\mu)$ 
universality and quote branching fractions
for $D \to X e^+ \nu_e$, which are better known than the corresponding
semimuonic values.  With \cite{PDG} $\b(D^0 \to X e \nu) = (6.49 \pm 0.11)\%$
and $\tau(D^0) = (410.1 \pm 1.5)$ fs, we have $\Gamma(D^0 \to X e \nu) =
(1.583 \pm 0.027) \times 10^{11}$ s$^{-1}$, while with $\b(D^+ \to X e^+ \nu_e
= (16.07 \pm 0.30)\%$ and $\tau(D^+) = (1040 \pm 7)$ fs, we have $\Gamma
(D^+ \to X e^+ \nu_e) = (1.545 \pm 0.031) \times 10^{11}$ s$^{-1}$.  The
values for $D^0$ and $D^+$ are equal to better than $1 \sigma$.  Averaging them
we obtain $\bar \Gamma(D \to X \ell^+ \nu_\ell) = (1.567 \pm 0.020) \times
10^{11}$ s$^{-1}$. We have ignored a possible difference between the small
Cabibbo-suppressed semileptonic decay rates of $D^0$ and $D^+$. 

The $D_s$ semileptonic decay rate is significantly smaller than that of the
non-strange $D$ mesons.  With $\b(D_s^+ \to X e^+ \nu_e) = (6.5 \pm 0.4)\%$ and
$\tau(D_s^+) = (500 \pm 7)$ fs, we have $\Gamma(D_s^+ \to X e^+ \nu_e) =
(1.300 \pm 0.082) \times 10^{11}$ s$^{-1}$ \cite{pred}, or
\beq \label{eqn:cr}
\frac{\Gamma(D_s^+ \to X \ell^+ \nu_\ell)}{\bar \Gamma(D \to X \ell^+
\nu_\ell)} = 0.830 \pm 0.053~.
\eeq
Thus, the semileptonic $D_s$ decay rate is $(17.0 \pm 5.3)\%$ lower than
the charge-averaged non-strange $D$ semileptonic rate.  We now show that the
model described in Sec.\ \ref{sec:eq} reproduces this inequality.
We will first study Cabibbo-favored decays and then calculate small corrections
from Cabibbo-suppressed decays. 

The semileptonic decays of $D$ are assumed to be dominated by 
Cabibbo-favored $K \ell \nu$ and $K^* \ell \nu$ in the ratio of $1:3$.
For the $D_s$ semileptonic decay we
need to know the $s \bar s$ content of the $\eta$ and $\eta'$, the two mesons
we are assuming dominate the pseudoscalar final state.  We shall quote results
for two extremes of octet-singlet mixing angles $\theta_\eta$ seen in the
literature \cite{Isgur:1975ib,etamix}.  These are summarized in
Table \ref{tab:mix}, where
\beq
\eta = c^{(\eta)}_n(u \bar u + d\bar d) + c^{(\eta)}_s s \bar s~,~~
\eta' = c^{(\eta')}_n(u \bar u + d\bar d) + c^{(\eta')}_s s \bar s~.
\eeq

\begin{table}
\caption{Octet-singlet mixing assumptions for $\eta$ and $\eta'$
\label{tab:mix}}
\begin{center}
\begin{tabular}{c c c c c} \hline \hline
$\theta_\eta$ & $ c^{(\eta)}_n$ & $c^{(\eta)}_s$ & $c^{(\eta')}_n$ &
 $c^{(\eta')}_s$ \\ \hline
$9.74^\circ$ & $\frac{1}{2}$ & $- \frac{1}{\s}$ & $\frac{1}{2}$ &
 $\frac{1}{\s}$ \\
$19.47^\circ$ & $\frac{1}{\st}$ & $- \frac{1}{\st}$ & $\frac{1}{\sx}$ &
$\frac{2}{\sx}$ \\ \hline \hline
\end{tabular}
\end{center}
\end{table}

The mixing angle $\theta_\eta = 9.74^\circ$ was proposed by Isgur
\cite{Isgur:1975ib}, while the quark composition corresponding to $\theta_\eta =
19.47^\circ$ has been proposed by several authors \cite{etamix} on
phenomenological grounds.

With the mixing angle $\theta_\eta = 9.74^\circ$, the $\eta$ and $\eta'$ each
consist half of nonstrange and half of strange quarks.  The weighted average
of $f$ for $D_s$ decays is then
\beq
\bar f(D_s) = \frac{1}{4} \left[ \frac{1}{2}(0.5682) + \frac{1}{2}(0.1781)
\right] + \frac{3}{4} (0.1395) = 0.1979~~(\theta_\eta = 9.74^\circ)~.
\eeq
One then predicts
\beq
\frac{\Gamma(D_s \to X \ell \nu)}{\bar \Gamma(D \to X \ell \nu)} = \left(
\frac{M(D_s)}{\bar M(D)} \right)^5 \frac{\bar f(D_s)}{\bar f(D)} =
 (1.3022)(0.6805) = 0.886~~(\theta_\eta = 9.74^\circ)~.
\eeq

With the mixing angle $\theta_\eta = 19.47^\circ$, the $\eta$ is composed
$1/3$ of strange and 2/3 of nonstrange quarks, while the $\eta'$ is
2/3 strange and 1/3 nonstrange.  The corresponding values of $\bar f(D_s)$
and the semileptonic rate ratio are
\beq
\bar f(D_s) = \frac{1}{4} \left[ \frac{1}{3}(0.5682) + \frac{2}{3}(0.1781) 
\right] + \frac{3}{4} (0.1395) = 0.1817~~(\theta_\eta = 19.47^\circ)~,
\eeq
\beq
\frac{\Gamma(D_s \to X \ell \nu)}{\bar \Gamma(D \to X \ell \nu)} = \left(
\frac{M(D_s)}{\bar M(D)} \right)^5 \frac{\bar f(D_s)}{\bar f(D)} = (1.3022)(0.6246)
= 0.813~~(\theta_\eta = 19.47^\circ)~.
\eeq
Both ratios are compatible with the experimental one.
%
\begin{table}
\caption{Ratios $R$ of $D_s$ semileptonic branching fractions to $\b(D_s \to
\phi \ell \nu_\ell)$.
\label{tab:comp}}
\begin{center}
\begin{tabular}{c c c c} \hline \hline
 & $R(\eta \ell \nu_\ell)$ & $R(\eta' \ell \nu_\ell)$ & $R(\eta')/R(\eta)$ \\
\hline
$\theta_\eta = 19.47^\circ$ & 0.453 & 0.284 & 0.627 \\
$\theta_\eta = 9.74^\circ$ & 0.679 & 0.213 & 0.313 \\
Experiment & $1.07 \pm 0.13$ & $0.40 \pm 0.10$ & $0.37 \pm 0.09$ \\
\hline \hline
\end{tabular}
\end{center}
\end{table}

The experimental semileptonic branching fractions of $D_s$ are \cite{PDG}
\beq
\b(\eta \ell \nu_\ell) = (2.67 \pm 0.29)\%~,~~
\b(\eta' \ell \nu_\ell) = (0.99 \pm 0.23)\%~,~~
\b(\phi \ell \nu_\ell) = (2.49 \pm 0.14)\%~,
\eeq
The $\eta$ and $\eta'$ branching fractions are compared with each other,
with the $\phi$ branching fraction, and with predictions of the model,
in Table \ref{tab:comp}.  The pseudoscalar-to-vector ratio
is underestimated; nonetheless, the $\eta'/\eta$ ratio is compatible with
a value of $\theta_\eta$ toward the lower end of the range considered.
%
\begin{table}
\caption{Cabibbo-suppressed semileptonic decays of charmed mesons to 
lowest-lying pseudoscalar and vector states. Notations as
in Table \ref{tab:decays}.
\label{tab:CSdecays}}
\begin{center}
\begin{tabular}{c c c c c} \hline \hline
Decay & $M_i$ (MeV/$c^2$) & $M_f$ (MeV/$c^2$) & $x$ & $f(x)$ \\ \hline
$D^0 \to \pi^- \ell^+ \nu_\ell$ & 1864.83 & 139.57 & 0.00560 & 0.9571 \\
$D^0 \to \rho^- \ell^+ \nu_\ell$ & 1864.83 & 775.11 & 0.17276 & 0.2871 \\
$D^+ \to \pi^0 \ell^+ \nu_\ell$ & 1869.60 & 134.98 & 0.00521 & 0.9600 \\
$D^+ \to \eta \ell^+ \nu_\ell$ & 1869.60 & 547.85 & 0.08587 & 0.5353 \\
$D^+ \to \eta' \ell^+ \nu_\ell$ & 1869.60 & 957.78 & 0.26244 & 0.1460 \\
$D^+ \to \rho^0 \ell^+ \nu_\ell$ & 1869.60 & 775.49 & 0.17205 & 0.2886 \\
$D^+ \to \omega \ell^+ \nu_\ell$ & 1869.60 & 782.65 & 0.17524 & 0.2820 \\
$D^+_s \to K^0 \ell^+ \nu_\ell$ & 1968.47 & 497.61 & 0.06390 & 0.6256 \\
$D^+_s \to K^{*0} \ell^+  \nu_\ell$ & 1968.47 & 895.94 & 0.20716 &
 0.2227 \\ \hline \hline
\end{tabular}
\end{center}
\end{table}

The equality between semileptonic widths of $D^0$ and $D^+$ may be 
violated in $\Delta I=1/2$ Cabibbo-suppressed decays. 
We now study corrections from these decays to the total $D^0, D^+$ and 
$D^+_s$ semileptonic widths. Decay modes
into the lowest-lying pseudoscalar and vector mesons are listed
in Table \ref{tab:CSdecays} for the three charmed mesons. The weighted 
averages of the function $f$ for Cabibbo-suppressed decays of the three
mesons are: 
 \bea\label{fCS}
\bar f_{\rm CS}(D^0) & = & \frac{1}{4} f(D^0 \to \pi^-) + \frac{3}{4} f(D^0 \to \rho^-) 
= 0.4546~, \nonumber \\
\bar f_{\rm CS}(D^+) & = & \left\{ \begin{array}{c}
\frac{1}{4}\left[\frac{1}{2}f(D^+ \to \pi^0) + \frac{1}{4} f(D^+\to \eta) + 
\frac{1}{4}f(D^+\to\eta')\right]~~~~~~~~~~~~~~\cr
+ \frac{3}{4}\left[\frac{1}{2}f(D^+\to\rho^0) + 
\frac{1}{2}f(D^+\to\omega) \right]=0.3766~~~~~~(\theta_\eta=9.74^\circ)~,~~\cr
\frac{1}{4}\left[\frac{1}{2}f(D^+ \to \pi^0) + \frac{1}{3} f(D^+\to \eta) + 
\frac{1}{6}f(D^+\to\eta')\right]~~~~~~~~~~~~~ \cr
+ \frac{3}{4}\left[\frac{1}{2}f(D^+\to\rho^0) + 
\frac{1}{2}f(D^+\to\omega) \right]=0.3847~~~~~~(\theta_\eta=19.47^\circ)~, \cr
\end{array}\right. \nonumber \\
\bar f_{\rm CS}(D^+_s) & = & \frac{1}{4}f(D^+_s\to K^0) + 
\frac{3}{4}f(D^+_s \to K^{*0}) = 0.3234~. 
 \eea
 
 In order to calculate the difference between the total $D^0$ and
 $D^+$ semileptonic widths we include small differences between charged and 
 neutral meson masses, which affect also rates for Cabibbo-favored semileptonic 
 decays. Thus, instead of the charge-averaged value of $\bar f(D)$ in
 Eq.~(\ref{eq:fD}) 
 we now use two slightly different kinematic factors for $D^0$ and $D^+$,
 \bea\label{fD0+}
 \bar f(D^0)=\frac{1}{4}f(D^0\to K^-)+\frac{3}{4}f(D^0\to K^{*-})= 
 \frac{1}{4}(0.5987)+\frac{3}{4}(0.1895)=0.2918~,  \nonumber \\
 \bar f(D^+)=\frac{1}{4}f(D^+\to \bar K^0)+\frac{3}{4}f(D^+\to \bar K^{*0})= 
\frac{1}{4}(0.5955)+\frac{3}{4}(0.1880)=0.2899~.
 \eea

 Using the values in Eqs. (\ref{fCS}) and (\ref{fD0+}) we calculate
 \beq
 \frac{\Gamma(D^0\to X\ell\nu)}{\Gamma(D^+\to X\ell\nu)} = 
 \left[\frac{M(D^0)}{M(D^+)}\right]^5
 \frac{\bar f(D^0) + \tan^2\theta_c \bar  f_{\rm CS}(D^0)}
 {\bar f(D^+) + \tan^2\theta_c \bar  f_{\rm CS}(D^+)} = 
 \left\{ \begin{array}{c}
1.007~~(\theta_\eta=9.74^\circ)~\cr
1.005~(\theta_\eta=19.47^\circ)~, \cr
\end{array}\right.
 \eeq
 \beq
 \frac{\Gamma(D^+_s\to X\ell\nu)}{\Gamma(D^0\to X\ell\nu)} = 
 \left[\frac{M(D^+_s)}{M(D^0)}\right]^5
 \frac{\bar f(D^+_s) + \tan^2\theta_c \bar  f_{\rm CS}(D^+_s)}
 {\bar f(D^0) + \tan^2\theta_c \bar  f_{\rm CS}(D^0)} = 
 \left\{ \begin{array}{c}
0.892~~(\theta_\eta=9.74^\circ)~\cr
0.825~(\theta_\eta=19.47^\circ)~, \cr
\end{array}\right.
 \eeq
 where~\cite{PDG} $\tan\theta_c=|V_{us}/V_{ud}|=0.231$.
 Thus, in our model the $D^0$ and $D^+$ total semileptonic widths are 
 predicted to be equal within less than one percent 
 independent of the $\eta-\eta'$ mixing angle.
 The observed ratio [see the discussion before Eq.\ (3)] is
\beq
\frac{\Gamma(D^0\to X\ell\nu)}{\Gamma(D^+\to X\ell\nu)} = \frac{1.583 \pm
0.027}{1.545 \pm 0.031} = 1.025 \pm 0.027~,
\eeq
so the errors on branching fractions need to be improved considerably before
the predicted deviation of this ratio from 1 can be identified.  The predicted
ratio of $D_s$ and $D$ total semileptonic widths, which depends somewhat 
on the mixing angle, is consistent with the experimental value (\ref{eqn:cr})
within about $1\sigma$.  An alternative interpretation of the semileptonic
width difference observed for $D_s$ and $D$ mesons in terms of a weak
annihilation amplitude contributing to $D_s$ decays
\cite{Voloshin:2001xi,Bigi:2009ym,Ligeti:2010vd} has been shown to be 
disfavored by the measured lepton energy spectrum in
these decays~\cite{Gambino:2010jz}.
  
A similar discussion may be applied to $\Lambda_c$ semileptonic decays.
Here we will neglect Cabibbo-suppressed decays in view of the large
experimental uncertainty in the observed semileptonic decay rate.
Only one early value has been published \cite{Vella:1982ei}: $\b(\Lambda_c \to
e^+ X) = (4.5 \pm 1.7) \%$ (see also Ref.~\cite{Albrecht:1991bu}),
leading when combined with the $\Lambda_c$ lifetime
\cite{PDG} of $(200 \pm 6)$ fs to
\beq
\Gamma(\Lambda_c \to e^+ X) = (2.25 \pm 0.85) \times 10^{11}~{\rm s}^{-1}~.
\eeq
Comparing this with the charge-averaged nonstrange $D$ semileptonic decay
rate $\bar \Gamma(D \to e^+ X)=(1.567 \pm 0.020) \times 10^{11}~{\rm s}^{-1}$,
we have
\beq
\frac{\Gamma(\Lambda_c \to e^+ X)}{\bar \Gamma(D \to e^+ X)} = 1.44 \pm 0.54~,
\eeq
poorly determined but with a central value considerably above 1.
The model predicts this ratio to be
\beq
\frac{\Gamma(\Lambda_c \to e^+ X)}{\bar \Gamma(D \to e^+ X)} = \left[
\frac{M(\Lambda_c)}{\bar M(D)} \right]^5 \left[ \frac{f(\Lambda_c)}
{\bar f(D)} \right] = (2.753)(0.606) = 1.67~.
\eeq
This prediction should be compared with an estimate of about 1.2 based on a 
heavy quark expansion including $1/m^2_c$ terms~\cite{Manohar:1993qn}.
Reducing the experimental error in ${\cal B} (\Lambda_c \to e^+ X)$ by a factor
of three would be a useful first step in testing these predictions.

\section{Application to beauty decays \label{sec:b}}

\begin{table}
\caption{Semileptonic branching fractions, total lifetimes, and
semileptonic decay rates of $B$ and $B_s$ mesons.
\label{tab:bdec}}
\begin{center}
\begin{tabular}{c c c c} \hline \hline
Meson & $\b_{SL}$ &  Lifetime   & $\Gamma_{SL}$ (units \\
      &    (\%)   & $\tau$ (fs) & of $10^{10}$ s) \\ \hline
$B^0$ & $10.33\pm0.28$ & $1525\pm9$ & $6.77\pm0.19$ \\ 
$B^+$ & $10.99\pm 0.28$ & $1638\pm11$ & $6.71\pm0.18$ \\
$B_s$ & $7.9\pm2.4$ & $1472^{+24}_{-26}$ & $5.4\pm1.6$ \\ \hline \hline
\end{tabular}
\end{center}
\end{table}

The experimental semileptonic branching fractions for beauty decays
and the corresponding decay rates \cite{PDG} are summarized in Table
\ref{tab:bdec}.  For $B$ decays we quote the $X \ell \nu$ branching fractions
based on assuming $e$--$\mu$ universality, while for $B_s$ decays we
quote the branching fraction to $D_s^- \ell^+ \nu_\ell X$, which is the
only one available in Ref.\ \cite{PDG}.  The semileptonic
nonstrange $B$ decay rates are consistent with one another within better
than $1 \sigma$, while the 30\% error on the $B_s$ semileptonic decay rate 
prevents one from making any crisp statement about its ratio to the nonstrange
rates.  To predict this ratio,
we proceed as we did for charm decays, evaluating the weighted averages
of the function $f$ for CKM-favored $B$ and $B_s$ decays:
\bea
\bar f(B) = &\frac{1}{4} f(B \to D) + \frac{3}{4} f(B \to D^*) = 
 \frac{1}{4}(0.4050) + \frac{3}{4} (0.3518) = 0.3651~,~~\\
\bar f(B_s) = &\frac{1}{4} f(B_s \to D_s) + \frac{3}{4} f(B_s \to D_s^*) =
 \frac{1}{4}(0.3785) + \frac{3}{4}(0.3268) = 0.3398~.
\eea
We then predict
\beq\label{eq:Bs/B}
\frac{\Gamma(B_s \to X \ell^+ \nu_\ell)}{\bar \Gamma(B \to X \ell \nu_\ell)}
= \left( \frac{M(B_s)}{\bar M(B)} \right)^5 \frac{\bar f(B_s)}{\bar f(B)}
= (1.0851)(0.9306) = 1.010~.
\eeq
We are thus led to expect an {\it enhancement} by one percent of the
the ratio of the strange to nonstrange $B$ semileptonic decay rates.  It
will be interesting to see if this prediction can be tested in forthcoming
experiments at lepton or hadron colliders.

\begin{table}
\caption{CKM-suppressed semileptonic decays of beauty mesons to 
lowest-lying pseudoscalar and vector states. Notations as  in Table \ref{tab:decays}.
\label{tab:CKMb}}
\begin{center}
\begin{tabular}{c c c c c} \hline \hline
Decay & $M_i$ (MeV/$c^2$) & $M_f$ (MeV/$c^2$) & $x$ & $f(x)$ \\ \hline
$B^0 \to \pi^- \ell^+ \nu_\ell$ & 5279.50 & 139.57 & 0.000699& 0.9945 \\
$B^0 \to \rho^- \ell^+ \nu_\ell$ & 5279.50 & 775.11 & 0.021555 & 0.8490 \\
$B^+ \to \pi^0 \ell^+ \nu_\ell$ & 5279.17 & 134.98 & 0.000654 & 0.9948 \\
$B^+ \to \eta \ell^+ \nu_\ell$ & 5279.17 & 547.85 & 0.010769 & 0.9202 \\
$B^+ \to \eta' \ell^+ \nu_\ell$ & 5279.17 & 957.78 & 0.032915 &  0.7813\\
$B^+ \to \rho^0 \ell^+ \nu_\ell$ & 5279.17 & 775.49 & 0.021579 & 0.8489 \\
$B^+ \to \omega \ell^+ \nu_\ell$ & 5279.17 & 782.65 & 0.021979 &  0.8464\\
$B_s \to K^- \ell^+ \nu_\ell$ & 5366.3 & 493.68 & 0.008463 &  0.9364\\
$B_s \to K^{*-} \ell^+  \nu_\ell$ & 5366.3 & 891.66 & 0.027609 & 0.8121\\
\hline \hline
\end{tabular}
\end{center}
\end{table}

Corrections from CKM-suppressed decays to the ratio (\ref{eq:Bs/B}) and to the 
equality of $B^0$ and $B^+$ semileptonic widths  are expected to be very small
as they are proportional to $(|V_{ub}|/|V_{cb}|)^2 \simeq 0.01$.
The two ratios of total inclusive widths are given by 
 \beq
 \frac{\Gamma(B^0\to X\ell\nu)}{\Gamma(B^+\to X\ell\nu)} = 
 \left[\frac{M(B^0)}{M(B^+)}\right]^5
 \frac{\bar f(B^0) + |V_{ub}/V_{cb}|^2 \bar  f_{\rm CKMS}(B^0)}
 {\bar f(B^+) + |V_{ub}/V_{cb}|^2 \bar  f_{\rm CKMS}(B^+)} 
  \eeq
 \beq
 \frac{\Gamma(B_s\to X\ell\nu)}{\Gamma(B^0\to X\ell\nu)} = 
 \left[\frac{M(B_s)}{M(B^0)}\right]^5
 \frac{\bar f(B_s) + |V_{ub}/V_{cb}|^2 \bar  f_{\rm CKMS}(B_s)}
 {\bar f(B^0) + |V_{ub}/V_{cb}|^2 \bar  f_{\rm CKMS}(B^0)} 
 \eeq
The weighted averages of the functions $f$ for CKM-suppressed decays 
are denoted $\bar f_{\rm CKMS}$.
For completeness, we calculate these functions using decay modes
listed in Table \ref{tab:CKMb}:
 \bea\label{CKMb}
\bar f_{\rm CKMS}(B^0) & = & \frac{1}{4} f(B^0 \to \pi^-) + \frac{3}{4} f(B^0 \to \rho^-) 
= 0.8854~, \nonumber \\
\bar f_{\rm CKMS}(B^+) & = & \left\{ \begin{array}{c}
\frac{1}{4}\left[\frac{1}{2}f(B^+ \to \pi^0) + \frac{1}{4} f(B^+\to \eta) + 
\frac{1}{4}f(B^+\to\eta')\right]~~~~~~~~~~~~~~\cr
+ \frac{3}{4}\left[\frac{1}{2}f(B^+\to\rho^0) + 
\frac{1}{2}f(B^+\to\omega) \right]=0.8664~~~~~~(\theta_\eta=9.74^\circ)~,~~\cr
\frac{1}{4}\left[\frac{1}{2}f(B^+ \to \pi^0) + \frac{1}{3} f(B^+\to \eta) + 
\frac{1}{6}f(B^+\to\eta')\right]~~~~~~~~~~~~~ \cr
+ \frac{3}{4}\left[\frac{1}{2}f(B^+\to\rho^0) + 
\frac{1}{2}f(B^+\to\omega) \right]=0.8693~~~~~~(\theta_\eta=19.47^\circ)~, \cr
\end{array}\right. \nonumber \\
\bar f_{\rm CKMS}(B_s) & = & \frac{1}{4}f(B_s\to K^-) + 
\frac{3}{4}f(B_s \to K^{*-}) = 0.8432~. 
 \eea
 
The approximate equality of the three values of $\bar f_{\rm CKMS}$ multiplying 
$|V_{ub}/V_{cb}|^2$ imply their negligible effect on the above two ratios of
widths.  This applies also to the factor $[M(B^0)/M(B^+)]^5 = 1.0003$ entering
the first ratio.  The largest correction to this ratio, a few parts in a
thousand, comes from $\bar f(B^0)/\bar f(B^+)$. Using masses for charged and
neutral $B, D$ and $D^*$ mesons~\cite{PDG} we find:
\bea
  \frac{\Gamma(B^0\to X\ell\nu)}{\Gamma(B^+\to X\ell\nu)} &\approx &
 \frac{\bar f(B^0)}{\bar f(B^+)} = \frac{0.3644}{0.3657} = 0.996~,\\
  \frac{\Gamma(B_s\to X\ell\nu)}{\Gamma(B^0\to X\ell\nu)} & \approx & 
 \left[\frac{M(B_s)}{M(B^0)}\right]^5 \frac{\bar f(B_s)}{\bar f(B^0)} = 1.012~.
\eea

We now turn to the semileptonic decay rate of $\Lambda_b$.  No inclusive
semileptonic decay branching fraction is quoted in Ref.\ \cite{PDG}.  
A similar method to the one discussed for $\Lambda_c$ lead to the prediction
\beq\label{Lambda_b}
\frac{\Gamma(\Lambda_b \to X \ell^- \bar \nu_\ell)}{\bar \Gamma(B \to X \ell^+
\nu_\ell)}=\left[ \frac{M(\Lambda_b)}{\bar M(B)} \right]^5 \frac{f(\Lambda_b)}{\bar f(B)}
 = (1.367)(0.829)=1.134~,
\eeq
where $\bar \Gamma$ is the average of charged and neutral $B$ decay rates.
This represents a considerable departure from the expectation of an approach
using operator product and heavy quark expansions which predicts this ratio to be
1.03~\cite{Manohar:1993qn}.
(See also Ref.~\cite{Jin:1997in} where this ratio is calculated to be
around 1.05.) A departure from unity similar to (\ref{Lambda_b}) is seen when
comparing total decay rates as quoted in Ref.~\cite{PDG},
\beq
\frac{\Gamma(\Lambda_b)}{\bar \Gamma(B)} = \frac{\bar \tau(B)}{\tau(\Lambda_b)}
= \frac{(1.570 \pm 0.007)~{\rm ps}}{(1.391^{+0.038}_{-0.037})~{\rm ps}} = 1.129
\pm 0.031~.
\eeq
A somewhat smaller value~\cite{al.:2010pj}, $\Gamma(\Lambda_b)/\bar\Gamma(B) 
= 1.024 \pm 0.032$, was reported while we were completing the writeup of this
paper.
 
\section{Summary \label{sec:sum}}

We have presented a simplified model for estimating ratios of semileptonic
decay rates of hadrons containing charm and bottom quarks.  The model uses
kinematic factors appropriate to free-fermion decays, but endows the initial
and final fermions with the physical masses of ground-state hadrons.  This
approach may be thought of as a cartoon version of local quark-hadron duality.
It appears to reproduce known ratios of rates for charm decays, including the
suppression of the  $D_s$ semileptonic rate by $(17.0 \pm 5.3)\%$ relative
to those of the non strange $D^0$ and $D^+$ (equal within errors).  For
hadrons containing $b$ quarks, it predicts an enhancement of about 1.2\%
for $\Gamma(B_s \to X \ell \nu)$ and about 13\% for $\Gamma(\Lambda_b \to
X \ell \nu)$ relative to $\bar\Gamma(B \to X \ell \nu)$.  The latter result
represents a significant deviation from expectations of the operator product 
and heavy quark expansion, and is similar to the departure from unity
exhibited by the ratio of total decay rates.

The prospects for testing differences in semileptonic decay rates for $B$ and
$B_s$ at the 1--2\% level are challenging.  The best chance we see would
involve the use of tagged $B_s$ decays, such as obtained by the Belle
Collaboration in a large sample of $B_s$--$\bar B_s$ pairs.  It may be more
feasible to study the considerably larger deviation from unity predicted for
the ratio of baryon to meson semileptonic decay rates.  Given the large sample
of charmed baryons produced at $B$ factories and foreseen at LHCb, it would
also be helpful to perform an improved measurement of ${\cal B}(\Lambda_c \to
e^+ X)$, to check our prediction that the corresponding inclusive semileptonic
decay rate is 1.67 times the average for non-strange $D$ mesons.  We look
forward to such tests and to a first measurement of ${\cal B}(\Lambda_b \to
e^-X)$.

\section*{Acknowledgments}

We thank Sheldon Stone for asking the question that led to this investigation
and Martin Beneke, Jernej Kamenik, Maxim Khlopov, Ulrich Nierste, and Sheldon
Stone for useful communications.
This work was supported in part by the United States Department of Energy
through Grant No.\ DE FG02 90ER40560 (JR).

\end{document}